\documentstyle[aps]{revtex}
\tightenlines
\begin{document}                                                 
\rightline{\bf Euro. Phys. J. A4 (1999) 217-220}
\begin{center}
\Large Excitation Energy as a Basic Variable to Control Nuclear Disassembly
\end{center}

\begin{center}
{\ Y. G.\ MA$^{a,b,c}$\footnotemark
\footnotetext{\footnotesize Corresponding Author. Email: mayugang@public1.sta.net.cn},
 \ Q. M.\ SU$^b$, \ W. Q.\ SHEN\ $^{b}$,
 \ J. S.\ WANG$^b$, \\ D. Q.\ FANG$^b$, \ X. Z.\ CAI$^b$, \ H. Y.\ ZHANG$^b$,
 \ D. D. HAN$^d$}
\vspace{.5cm}

$^{a}$CCAST (World Laboratory), P. O. Box 8730, Beijing 100080 , China

$^{b}$Shanghai Institute of Nuclear Research, Chinese Academy of Sciences,
Shanghai 201800, China$\footnotemark
\footnotetext{\footnotesize Mailing address}$

$^{c}$Fudan - T. D. Lee Physics Laboratory, Fudan University, Shanghai 200433
,China

$^{d}$East China Normal University, Shanghai 200053, China
\end{center}

\begin{abstract}
Thermodynamical features of Xe system is investigated as  functions of
temperature and freeze-out density
in the frame of lattice gas model. The calculation shows
different  temperature dependence of physical observables at different
freeze-out density. In this case, the critical temperature when the phase
transition takes place depends on the freeze-out density. 
However, a unique critical excitation energy 
reveals regardless of freeze-out density
when the excitation energy is used as a variable insteading of temperature.
Moreover, the different behavior of other physical observables with temperature
due to different  $\rho_f$ vanishes when excitation energy replaces temperature.
It indicates that the excitation energy can be seen as a more basic quantity
to control nuclear disassembly.
\end{abstract}
\vspace{0.5cm}


The phase transition and critical phenomenon of small systems is an
interesting subject 
in recent nuclear physics research.
The break-up of nuclei due to violent collisions
into several intermediate mass fragments (IMF), can be viewed as critical
phenomenon as observed in fluid, atomic, and other
systems. It prompts the possible signature on the liquid gas
phase transition  of the nuclear system. On one hand, the onset 
of the multifragmentation \cite{Biza93}
and vaporization \cite{Rive96} channels    
can be seen as the signature of the boundaries of phase
mixture \cite{Gros90}. This is supported further by the fact that 
the caloric curve in a certain excitation energy range 
 \cite{Poch95} shows a saturate similar to a first order phase
transition, in the framework of statistical equilibrium models
\cite{Bond98}. On other hand, the observation of critical
exponents  parameters in the charged or mass
distribution of the
 multifragmentation system \cite{Haug96} can be interpreted as
an evidence of the phase transition. Recently, 
the lattice gas model (LGM) has been applied to treat phase
transition and critical phenomenon in 
the nuclear disassembly for isospin symmetrical \cite{Jpan95} 
and asymmetrical \cite{Sray97,Gulm98} nuclear systems.
LGM assumes a  freeze-out density $\rho_f$
with thermal equilibrium at temperature $T$.
The temperature was adopted naturally as a variable to study the feature of
disassembly in nearly all previous calculation of LGM.
In this paper, we will illustrate that the excitation energy can be
taken as a more basic quantity to control the disassembly of nuclear
system rather than temperature
via studying the features of critical phenomenon and other physical
observables  in   the lattice gas model.

In the lattice gas model, $A$ nucleons with an
occupation number $s$ which is defined as $s$ = 1 (-1) for a proton 
(neutron) or $s$ = 0 for a vacancy, are placed in the $L$ sites of 
lattice. Nucleons in the nearest neighbouring sites have 
interaction with an energy $\epsilon_{s_i s_j}$. The hamiltonian
 is written by 
$ E = \sum_{i=1}^{A} \frac{P_i^2}{2m} - \sum_{i < j} \epsilon_{s_i s_j}s_i s_j .  $
The interaction constant $\epsilon_{s_i s_j}$ is related  to
 the binding energy of the nuclei. Here $\epsilon_{nn,pp}$
 = $\epsilon_{-1-1,11}$ = 0 MeV, $\epsilon_{pn,np}$ 
 = $\epsilon_{1-1,-11}$  = - 5.33 MeV is used. 
The freeze-out density of disassembling system is
$\rho_f$ = $\frac{A}{L} \rho_0$ where $\rho_0$ is the normal nucleon
density. The disassembly of the system is to be calulated 
at $\rho_f$, beyond which nucleons are too far apart to interact.
$N + Z$ nucleons are put in $L$ cubes with size $l^3$ by Monte Carlo sampling using the
Metropolis algorithm. 
Once the nucleons have been placed, their momentum is generated by a Monte 
Carlo sampling of Maxwell Boltzmann distribution. Various observables
can be calculated in a straightforward fashion.

One of the basic measurable quantities is the distribution of fragment mass. In this LGM, two neighboring nucleons  are viewed to be in the same fragment if their relative kinetic energy 
is insufficient to overcome the attractive bond: 
$P_r^2/2\mu + \epsilon_{np} < 0 $. 
Once  the fragment mass distribution is built, we can extract the effective
power law parameters via fitting the mass  distribution of
fragments  with
$Y(A_i) \propto A_i^{-\tau}$
and its second moment  of fragment distribution defined as \cite{Camp88}
$S2 = \frac{ \sum_{i \neq Amax} {A_i^2*n_i(A_i)}}{A}$,         
where $n_i(A_i)$ is the number of fragments with $A_i$ nucleons and the 
sum excludes the largest cluster $A_{max}$. 
There are a minimum of $\tau$ and a maximum of S2 at critical point
for an infinite system. Besides the above quantities, we will use
the average multiplicity  $<IMF>$ of IMF  and the information entropy $H$ to
search the critical point \cite{Mayg95,Mayg98}.
$H$ was defined firstly by Shannon in information theory \cite{Denb85} and
can be introduced into nuclear dissociation \cite{Mayg98}, it reads
$ H = -\sum_{i} {p_i * ln(p_i)} $,
where $p_i$ is the probability having $"i"$ produced particles in each event,
the  sum is taken over all multiplicities of products from the
disassembling system.
$H$ reflects the capacity of the information or the extent of disorder.
  
We choose the medium size nuclei $^{129}$Xe as an example to analyze the
nuclear disassembly. 
Three freeze-out densities of 0.18$\rho_0$,
0.38$\rho_0$, and 0.60$\rho_0$, corresponding to the lattice
size of 9$^3$, 7$^3$ and 6$^3$ respectively , were used.
The calculations were  performed from 3 MeV to 7 MeV and 1000 events were
accumulated at each  temperature and freeze-out density.

We  show the temperature and freeze-out density dependences of $\tau$,
$<IMF>$, $H$ and $S2$ in the left columu of Fig.1. First, the critical temperatures 
determined by the extreme values are the same for the same freeze-out density, indicating that critical phenomenon really exsits. Second, the freeze-out density determines the temperature where the critical point is reached in this model. The critical point takes place at 4.25 MeV for freeze-out density  0.18 $\rho_0$, 5.5 MeV for 0.38 $\rho_0$, and 6.5 MeV for 0.60 $\rho_0$, respectively. Obviously, the higher the freeze-out density, the higher the critical
temperature. In this case, the critical behavior is determined by the two variables, namely temperature $T$ and freeze-out density $\rho_f$ as observed in Pan and Das Gupta \cite{Jpan95}.
But the extraction of $T$ and $\rho_f$ is difficulty in view of experiments. On one side, the different thermometers give different apparant temperatures \cite{Morr94,Mayg97}.
On other side, the freeze-out density is not directly measurable quantity in experiments. 
Moreover,  the different initial conditions of models maybe result in 
different freeze-out density during the confrontations of model calculations with experimental
data. All these facts complicate the accurate determination on temperature and freeze-out density and interfer with the correct extraction of physics.
So it is interesting to search other variables to locate the critical
phenomenon. A natural ideal is to adopt the excitation energy per
nucleon $E^*/A$, which can be deduced or reconstructed from the experiments,
especially in 4$\pi$ detector measurements \cite{Mayg97}.
In the lattice gas model it  can be  defined as
$ E^*/A = E_T - E_{g.s.} = \frac{3}{2}T + \epsilon_{n,p}\frac{N_{n,p}^T}{A} 
-\epsilon_{n,p} \frac{N_{n,p}^{g.s.}}{A},  $
where $N_{n,p}^T$ and $N_{n,p}^{g.s.}$ is the
number of the bonds of unlike nucleons at $T$ and in the ground state, respectively.
In a pure classical model the ground state corresponds to a cold nucleus at zero
 temperature and normal nuclear density where there is no kinetic energy and 
so that the ground state energy per nucleon is 
$-\epsilon_{n,p} \frac{N_{n,p}^{g.s.}}{A}$. Practically, $N_{n,p}^{g.s.}$ is determined by the
geometry and is equal to the maximum bond number of unlike nucleons possible for $^{129}Xe$. 
The Fig.1e shows the excitation energy per nucleon $E^*/A$ in different temperature and freeze-out density.  Noted that the curves of excitation energy are not linear with temperature.
If performing the differentials for these curves, we can obtain an important
thermodymical quantity: the specific heat per nucleon at constant volume (or density) as 
 $C_v/A = \frac{\partial (E^*/A)}{\partial T}$. The Fig.1f shows $C_v/A$ as a function of  $T$ for the disassembling system $^{129}Xe$ at three $\rho_f$. Clearly, the peaks of $C_v/A$ exist for the systems at each $\rho_f$ and locate closely at theirself critical temperatures, which supports strongly the viewpoint about critical feature as said above. By the mapping from $E^*/A$ to $T$, we replot the $\tau$, $<IMF>$, $H$, and $S2$ as a function of excitation energy instead of temperature in the right column of Fig.1. Now nearly all the critical points locate
at the same excitation energy regardless of the freeze-out density. Hence  the excitation energy can be viewed as a more basic quantity in controlling the reaction dissociation.

In order to illustrate this point further, Fig.2 gives the average mass of the largest fragment ($A_{max}$), the isotopic ratio R(p/d) between protons and deutrons, the isobaric
ratio R(t/$^3He$) between tritons and helium-3 and the ratio R(n/p) of emitted neutrons to protons as a function of temperature or excitation energy, respectively, in different freeze-out
density. Again, the discrepancies stemming from different freeze-out density minimize when the excitation energy is used as the variable, and the curves for different  $\rho_f$ merge approximately  into a single line.

To summarize, we studied the critical feature of Xe system and found
that the critical temperature where the phase transition occurs
changes with the freeze-out density  which
complicates the confrontation the theoretical predication with the experimental
results. Contrary, we found that a unique critical excitation energy
and the same excitation energy dependence of other physical observables
 reveals regardless of the freeze-out density, which indicates
that the excitation energy can be viewed as a basic parameter
to control the dissociation of the nuclear system. Unlike the
temperature and freeze-out density, the excitation energy can be well
extracted from experiments, especially for  experiments using
4$\pi$ multidetectors nowadays. Hence the use of excitation energy as a basic parameter
will make it easier and definite to extract
physics from the direct comparison between  the experimental data and the
theoretical calculation.

We thank Dr. J. Pan for providing the lattice gas code and fruitful discussion.
This work was supported  by the NSFC for Distinguished
Young Scholar under Grant No. 19725521, the
NSFC under Grant No. 19705012, the
Science and Technology Development Foundation of Shanghai under Grant No.
97QA14038, the Special Project of  the
Presidential Foundation of CAS,
and the Scientific Research Foundations for the Returned Overseas Chinese Scholars
by the National Human Resource Administration and Education Administration
of China.

\widetext
\begin{center}
Figure Captions
\end{center}

\figure{Fig.1: The  observables as a function of temperature
(the left columu) or excitation energy (the right columu)
 in different freeze-out density:
the $\tau$ parameter from the power law fit to mass distribution (a,g),
the average multiplicity of intermediate mass fragments $<IMF>$ (b,h),
the information entropy $H$ (c,i) and the second moment  S2 (d,j).
The mapping  from
temperature  to excitation energy is plotted in Fig.1e and the specific heat
is shown in Fig.1f.}

\figure{Fig.2: The average mass of the largest fragment in each event 
$A_{max}$ (a,e), the isotopic ratio R(p/d) between the protons and
deutrons (b,f),  the isobaric ratio R(t/$^3He$) between the 
tritons and the Helium-3 (c,g), and the ratio R(n/p) of neutrons
and protons (d,h). The left columu is ploted versus the
temperature $T$ and the right columu versus the excitation energy.}  


\begin{references}
\bibitem{Biza93}G. Bizard et al., Phys.\ Lett.\ B302 (1993) 162; A. Sch\"uttauf
et al., Nucl.\ Phys.\ A607 (1996) 457.
\bibitem{Rive96}M.F. Rivet et al., Phys.\ Lett.\ B388 (1996) 219;
B. Borderie et al., Phys.\ Lett.\ B388 (1996) 24.
\bibitem{Gros90}D.H.E. Gross et al., Rep.\ Prog.\ Phys.\ 53 (1990) 605 and
references therein.
\bibitem{Poch95}J. Pochodzalla et al., Phys.\ Rev.\ Lett.\ 75 (1995) 1040.
\bibitem{Bond98}J.P. Bondorf et al., Phys.\ Rev.\ C58 (1998) R27. 
\bibitem{Haug96}M.L. Gilkes et al., Phys.\ Rev.\ Lett.\ 73 (1994) 1590;
J.A. Hauger et al., Phys.\ Rev.\ Lett.\ 77 (1996) 235.
J.B. Elliot et al., Phys.\ Lett.\ B381 (1996) 35.
\bibitem{Jpan95}J. Pan and S. Das Gupta, Phys.\ Lett.\ B344 (1995) 29;
Phys.\ Rev.\ C51 (1995) 1384.
\bibitem{Sray97}S. Ray, J. Shamanna and T.T.S. Kuo, Phys.\ Lett.\ B392 (1997) 7.
\bibitem{Gulm98}F. Gulminelli and P. Chomaz, Preprint Lpcc 98-05.
\bibitem{Camp88}X. Campi et al., Phys.\ Lett.\ 208 (1988) 351.
\bibitem{Mayg95}Y.G. Ma et al., Phys.\ Rev.\ C51 (1995) 710.
\bibitem{Mayg98}Y.G. Ma et al., Chin.\ Phys.\ Lett. (accepted).
\bibitem{Denb85} K.G. Denbigh and J.S. Denbigh, Entropy in Relation to
Uncomplete Knowledge, Cambridge University Press, 1985.
\bibitem{Morr94}D.J. Morrissey et al., Annu. Rev. Nucl. Part. Sci.
 44 (1993) 676 and references therein.
\bibitem{Mayg97}Y.G. Ma et al., Phys.\ Lett.\ B 390 (1997) 41.


\end{references}
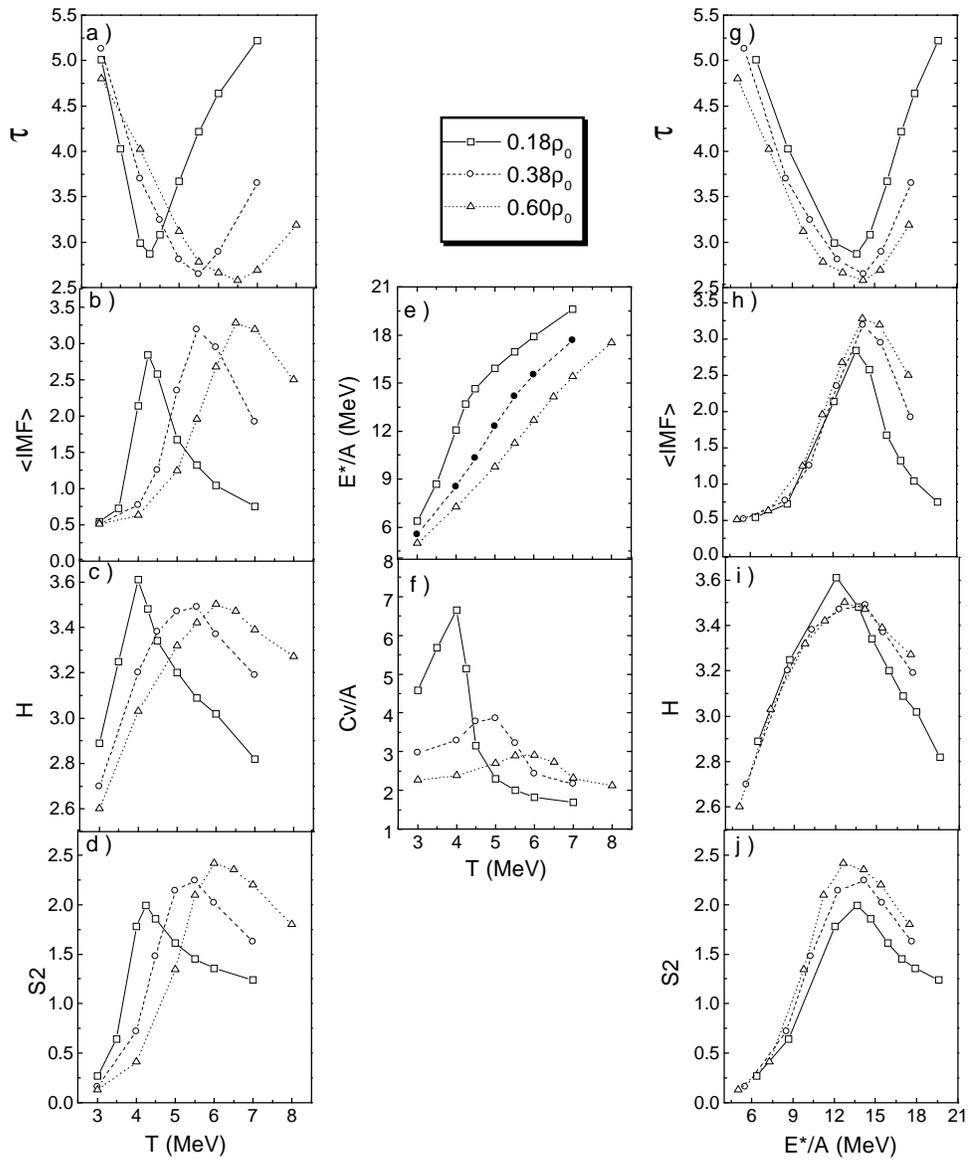
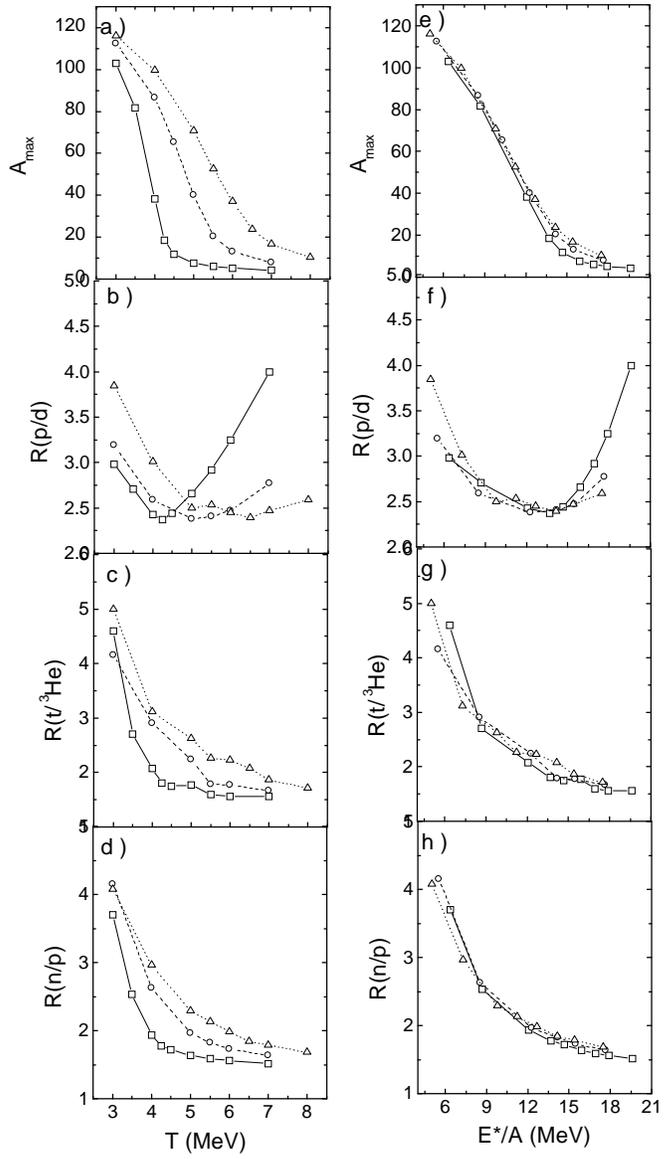
\end{document}